\documentclass[aps,pre,groupedaddress,twocolumn,showpacs]{revtex4}
\usepackage{amsmath}
\usepackage{amssymb}
\usepackage[final]{graphicx}

\begin{document}


\title{Coordinated and Uncoordinated Optimization of Networks}

\author{Markus Brede}
\affiliation{CSIRO Marine and Atmospheric Research, CSIRO Centre for Complex System Science, F C Pye Laboratory,
GPO Box 3023, Clunies Ross Street
Canberra ACT 2601, Australia}

\email{Markus.Brede@Csiro.au}


\date{\today}

\begin{abstract}
In this paper we consider spatial networks that realize a balance between an infrastructure cost (the cost of wire needed to connect the network in space) and communication efficiency, measured by average shortest pathlength. A global optimization procedure yields network topologies in which this balance is optimized. These are compared with network topologies generated by a competitive process in which each node strives to optimize its own cost-communication balance. Three phases are observed in  globally optimal configurations for different cost-communication trade-offs: (i) regular small worlds, (ii) star-like networks and (iii) trees with a centre of interconnected hubs. In the latter regime, i.e. for very expensive wire, power laws in the link length distributions $P(w)\propto w^{-\alpha}$ are found, which can be explained by a hierarchical organization of the networks. In contrast, in the local optimization process the presence of sharp transitions between different network regimes depends on the dimension of the underlying space. Whereas for $d=\infty$ sharp transitions between fully connected networks, regular small worlds and highly cliquish periphery-core networks are found, for $d=1$ sharp transitions are absent and the power law behaviour in the link length distribution persists over a much wider range of link cost parameters. The measured power law exponents are in agreement with the hypothesis that the locally optimized networks consist of multiple overlapping sub-optimal hierarchical trees.
\end{abstract}


\maketitle

\section{Introduction}
In recent years much interest has been focused on characterizing complex systems as networks. Seemingly universal characteristics have been revealed in analyses of many real-world systems such as power grids, the internet, metabolic and gene-regulatory networks and some social networks. Scale-free degree distributions \cite{Barabasi} and a small-world character are the prime examples \cite{Strogatz}. In the first case, many networks are found to have a degree distribution with a power law tail $p(k)\propto k^{-\gamma}$ with an exponent $\gamma$ in the range $2<\gamma<3$.  The second case, small worlds, describe networks that combine a high degree of cliquishness (reminiscent of, for example, an underlying spatial organisation) and small average shortest pathlength that scale logarithmically with the system size as in random graphs. Recent developments in the field are summarized in \cite{nets}.

It is important to find mechanisms that explain the observed network topologies of real-world systems. Many such mechanisms have been suggested to date, ranging from assembly mechnisms such as preferential attachment and varieties thereof  \cite{nets} to optimization \cite{Gopal,Sole}. Whereas the first approach assumes that a network is gradually assembled by a piecewise addition of new nodes according to certain stochastic rules, the latter approach assumes that the existing system represents the endpoint of an evolution that maximized some fitness measure. This idea was first advanced in \cite{Gopal}, which discussed mechanisms for the emergence of small-world networks. The authors of the study considered a set of nodes distributed in space and investigated networks that minimize a combination of pathlength and cost of wire needed to connect the nodes, while keeping the number of links fixed. The paper also argued that power laws in the link length distributions emerge from such an optimization principle, but no theoretical explanation was given.  This idea was extended in \cite{Sole}, which studied networks that optimize a trade-off between the cost of links and optimal communication measured by average shortest pathlength. If the first consideration dominates, optimal networks are star-like. If shortest pathlengths are more desirable than link economy, fully connected networks were found to be optimal. At a transition point between both regimes scale-free networks with an exponent $\alpha=3$ have been reported. Subsequent studies made use of the approach of \cite{Sole}, investigating optimal transport with congestion \cite{Banavar,Barth} or subject to targetted attacks \cite{Pongor}, synchronization \cite{Donetti,MB0,MB2} or networks that give rise to stable linear dynamics \cite{Variano,MB1}.

In this paper we revisit the results of \cite{Gopal} and \cite{Sole} and consider networks that realize a trade-off between cost of links and average shortest pathlength. However, unlike in \cite{Sole}, we do not measure the cost of links as simply proportional to the number of links. Instead, like \cite{Gopal} we investigate a set of nodes that are located in space and find the network that optimizes a trade-off between average shortest pathlength and the cost of the amount `wire' required to connect the nodes in space. In this way we introduce spatial constraints into the network optimization problem. Spatial constraints play an important role in the formation of various real-world networks. Likewise, pressures to minimize the network distance shape communication networks, for which the physical layout of the internet is an important example.

Since it appears more natural that links can be formed and removed as demanded by resource availability, we also do not fix the number of links in the optimization as it was done in \cite{Gopal}. Thus, in our model multiple short links can be transformed into one long link, which turns out to be of importance below. In this way the model is able to capture both the `connectivity' transition between sparse and dense networks discussed in \cite{Sole}, but also includes spatial constraints and a discussion of the spatial organization of the network arrangements becomes possible. In the first part of the study we undertake a detailed analysis of different regimes in the phase diagram of networks which realize varying trade-offs of cost of wire and communication.

All optimization of a global property (like pathlength and wire in this case) in a distributed system assumes (i) the functioning of the elementary components as a whole such that individual components benefit from an improved global performance and (ii) the possibility to coordinate alterations in system structure such that local and global improvements can be aligned, similar to conflicts between local and global optima that are the subject of competitive Game Theory. Relaxing these requirements, in the second part of the paper we also consider a localized optimization approach, where each node strives to optimize its own (local) trade-off between cost of wire and the average pathlength to other nodes. This concept of competitive optimization of a network by distributed agents appears again relevant. For instance, infrastructure and communication networks are not always formed according to a central planning scheme, but links are often established or removed when it appears profitable to a local actor.

Analyzing the differences between the globally and locally optimal solutions, we demonstrate that local optimization leads to a set of more `realistic' network topologies and a broadening of the parameter region where power laws in the link length distribution can be observed. Notably, the exponents of the power laws describing the link length distributions obtained from the competitive optimization process are much smaller than observed for globally optimal networks, thus being close to some exponents reported for various real-world systems.

Link length distributions of a power law form have been reported in several studies before and seem to occur in contexts as varied as for the internet \cite{Yook}, integrated circuits \cite{Payman}, the human cortex  \cite{Schuz} and certain regions of the human brain \cite{Egui}. For instance, for 2-dimensional systems Ref. \cite{Yook} reports an exponent of $-1$ for the internet and Ref. \cite{Payman} an exponent of $-.75$ for integrated circuits. Our study aims to contribute towards a better understanding of mechanisms that generate such power laws.

\section{The Global Optimum}

\label{Global}

\begin{figure*}[tbp]
\begin{center}
\includegraphics [width=.95\textwidth]{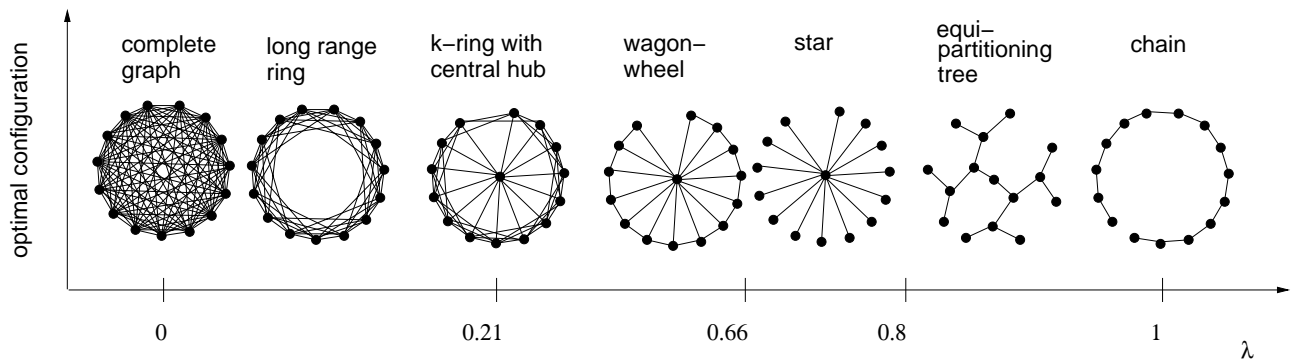}
\caption{Illustration of optimal networks in the order in which they appear when the trade-off parameter is increased from $\lambda=0$ to $\lambda=1$.}
\label{F-2}
\end{center}
\end{figure*}

\begin{figure}[tbp]
\begin{center}
\includegraphics [width=.45\textwidth]{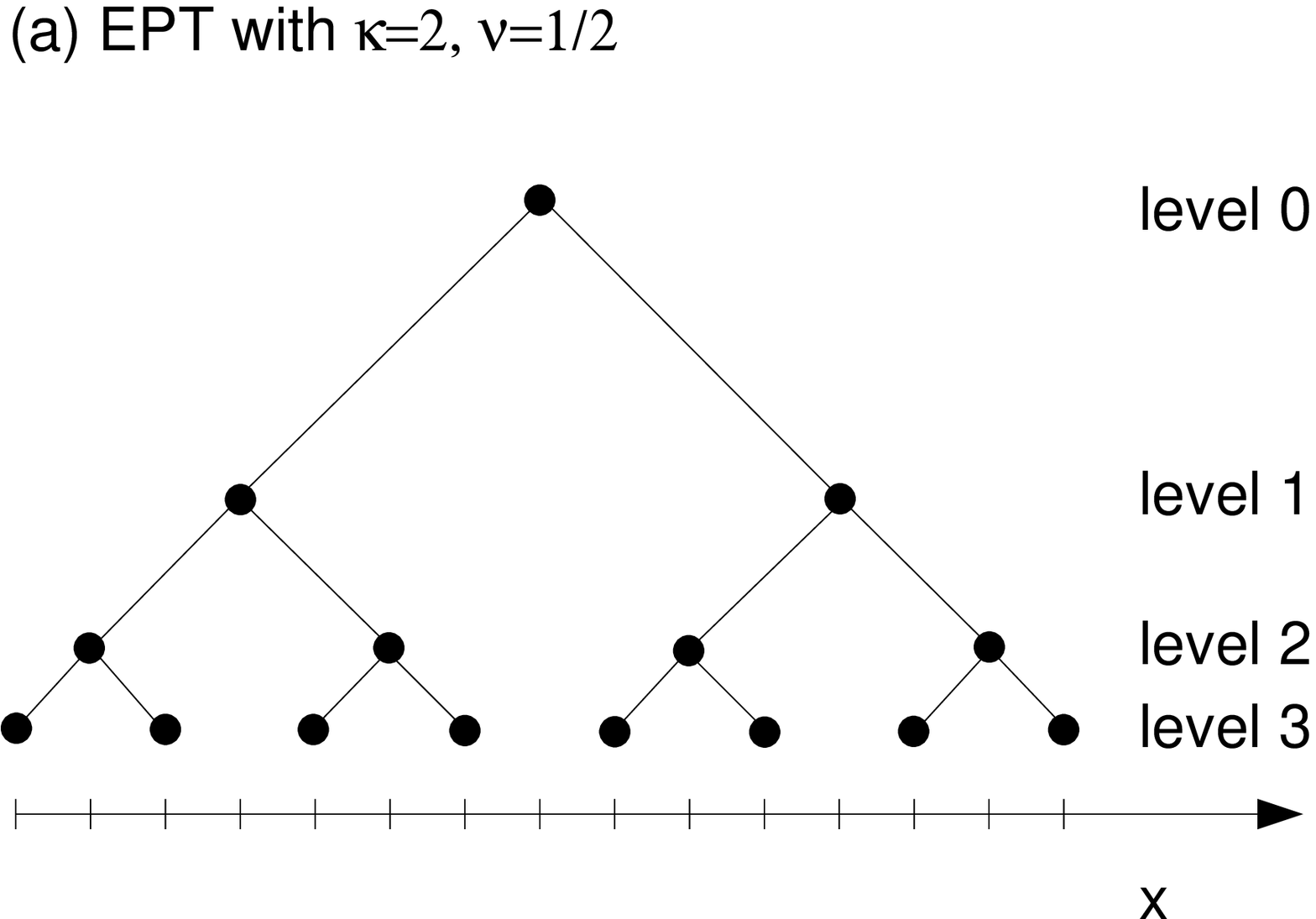}
\includegraphics [width=.45\textwidth]{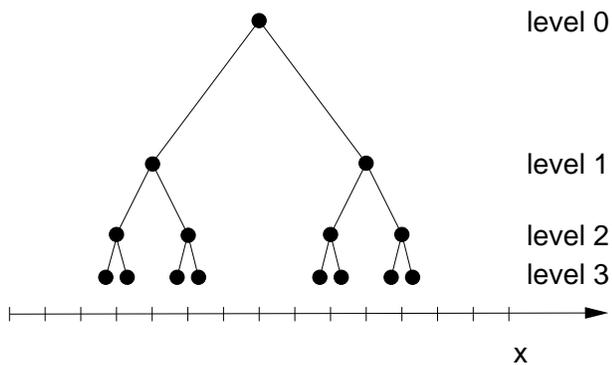}\\
\caption{Examples of equi-partitioning trees (EPT's) of $N=15$ nodes embedded in a 1d space without periodic boundary conditions: (a) an EPT with $\kappa=2$ and $\nu=1/2$, (b) an EPT with $\kappa=2$ and $\nu=2/3$. }
\label{F-3}
\end{center}
\end{figure}

Consider a set of $N$ nodes with labels $i=0,...,N-1$ and spatial positions $x_0,...,x_{N-1}$ and let $d(i,j)$ be the spatial distance between the nodes $i$ and $j$. The nodes are assumed to be uniformly distributed in a space with periodic boundary conditions, albeit the procedure we discuss below can easily be adapted to non-uniformly distributed nodes in space, if required in specific applications. For example in one dimension we have $x_i=\delta i$, $i=0,...,N-1$ and $d(i,j)=\min(|x_i-x_j|, N\delta-|x_i-x_j|)$ is a distance measure on that space, in which $\delta$ defines the spacing of the grid. Without loss of generality we assume $\delta=1$ for the following. Note, that this definition of a distance measure differs from that assumed in \cite{Gopal}, where the spatial distance between nodes on a 1d grid was defined by the spatial distance in 2d when the nodes are embedded on a circle. Advantages of our formulation are (i) that it appears more natural and does not require an embedding in a higher dimensional space and (ii) that its generalization into higher dimensions is obvious.

 In contrast to the spatial metric, the network distance between nodes $i$ and $j$, $l(i,j)$, is the minimum number of `hops' to get from $i$ to $j$ on the network. Thus, whereas the former distance measure is related to the cost of wire to establish connections between nodes, the latter is a measure for the efficiency of communication.

Generally, the amount of wire required to connect nodes in some d-dimensional space will depend on the dimension and topology of that space. Maximum distances in a d-dimensional space with periodic boundary conditions scale as $d_\text{max}=\sqrt{d}/2 N^{1/d}$, i.e. for fixed system size they decrease steeply with $d$. Since we mostly present numerical results (and the determination of average shortest pathlengths is numerically expensive) and are also interested in the interplay of many spatial scales in network formation, we restrict the bulk of the paper to $d=1$, where distances up to $d_\text{max}=N/2$ can be observed and only briefly comment on $d=2$ at the end of the paper.

Formally, we consider the problem to find network configurations $A=(a_{ij})_{i,j=0}^{N-1}$ that minimize an energy-like quantity $E(\lambda)$
\begin{align}
\label{E0}
 E(\lambda)=\lambda \overline{w}+(1-\lambda) \overline{l},
\end{align}
where $\overline{w}= \frac{2}{N(N-1)}\sum_{i<j} a_{ij} d(i,j)$ is the (normalized) amount of wire needed to realize the network $\Gamma_A$ given by its adjacency matrix $A$ and $\overline{l}=\frac{2}{N(N-1)}\sum_{i,j} l(i,j)$ is the average shortest pathlength on the network $\Gamma_A$. The parameter $\lambda$ is a measure for the relative weight of the cost of wire and desirability of efficient communication in Eq. (\ref{E0}).

To get a first handle on the problem, some limiting cases and particular network configurations are of interest, cf. also Fig. \ref{F-2} for some illustrations. We first discuss some simple limiting cases in subsection \ref{sub1} and then, in subsection \ref{sub2} we introduce and analyse equi-partitioning trees (EPT), a particular class of trees that have power law distributed link length distributions. This class of networks is important, because it establishes a link between power law link length distributions and the hierarchical structure of networks. Having considered these particular network configurations this then allows us to study transitions in the optimal topologies when the trade-off parameter $\lambda$ is varied in subsection \ref{sub3}. We conclude the section by the presentation and analysis of numerical results for optimal networks in subsection \ref{sub4}.

\subsection{Limiting cases: Complete graphs, stars and chains}
\label{sub1}

Some simple graph structures are of interest because they demarcate optimal solutions for the limiting situations $\lambda=0$ and $\lambda=1$. The first of these are complete graphs for which $\overline{l}=1$. Neglecting terms smaller than ${\cal O} (1)$ elementary calculations also yield $\overline{w}=N/4+1/4$, i.e.

\begin{align}
\label{Efull}
E_\text{full} (\lambda) = \lambda \left( N/4-3/4 \right) +1.
\end{align}
In passing we note that the link length distribution for the complete graph is uniform, i.e. all lengths occur with the same frequency.

The second limiting configuration of interest are star-like networks, in which one hub is connected to all other nodes. A back of the envelope calculation gives  $\overline{l}=2-2/N$ and $\overline{w}=1/2 (1+1/N)$ for such networks. Since we are interested in large networks we again neglect terms smaller than ${\cal O}(1)$ and obtain
\begin{align}
\label{Estar}
E_\text{star} (\lambda) = -3/2 \lambda + 2.
\end{align}
Note that the expression in Eq. (\ref{Estar}) is independent of the system size $N$. As for the complete graph the link length distribution for the star configuration is uniform.

Relevant as a limiting case are also regular grids, for instance a simple ring graph, for which only direct spatial neighbours are connected. Neglecting order ${\cal O}(1/N)$ terms one then has $\overline{l}=N/4+1/4$ while trivially $\overline{w}=2/N$. Hence
\begin{align}
\label{Ering}
E_\text{ring}=\lambda \left( 2/N-N/4-1/4 \right)+N/4+1/4.
\end{align}
The link length distribution of the ring is characterized by only one length scale which corresponds to the spacing of the grid.

\subsection{Equi-partitioning trees}
\label{sub2}

A further class of networks is of interest. These networks are hierarchical trees, which partition the underlying space into equal parts such that root nodes at a lower level of the hierarchy are always in the middle of the space occupied by the sub-trees at the higher levels of the hierarchy. Assuming that $N=2^{k+1}-1$ a recursive construction is the following:
\begin{itemize}
 \item [1.] Start with the interval $I_0=[0,2^{k+1}-1]$, i.e. the whole space.  The node in the middle, i.e. at position $2^{k}-1$ is the root node. Proceed with step 2 with both of the intervals $I_1=[0,2^{k}-1)$ and $I_2=(2^{k}-1,N]$ and connect the root node $2^k-1$ to the root nodes for both intervals.
 \item [2.] Given the nodes belonging to an interval $I=[n_1,n_2]$, if $n_1=n_2$ consider $m=n_1$ as the root node and stop the procedure. Otherwise,  select the node in the middle of $I$, $m$, which is assigned the role of root node. Proceed with step 2. for the intervals $I_1=[n_1,m)$ and $I_2=(m,n_2]$ and connect the root node $m$ to the root nodes of $I_1$ and $I_2$.
\end{itemize}
Figure \ref{F-3}a gives an illustration of such a tree and its embedding in the underlying space. If $N\not= 2^{k+1}-1$ the procedure can still be applied, resulting in imperfect trees that are not completely balanced. Above, we constructed trees by subdividing intervals into two equal parts, resulting in a graph where apart from the minimum and maximum level of the hierarchy all nodes have degree 3. Similarly, subdividing intervals into $\kappa$ parts will result in equi-partitioning trees with larger degree. These, however, have very similar properties to the trees constructed with $\kappa=2$ so that we only consider the case of $\kappa=2$ in detail below.

A simple calculation shows that for $N=2^{k+1}-1$ the total amount of wire used in the construction of an EPT with $\kappa=2$ is $k 2^k$ such that for large $N$ in leading order $\overline{w}=1/N \log_2 N$. Albeit a bit more elaborate (see appendix for details) the calculation of the average pathlength is also straightforward and yields
\begin{align}
 \overline{l}= 1/2 \log_2 N -1
\end{align}
for large $N$. Hence one has
\begin{align}
\label{Eept}
 E_\textrm{EPT}(\lambda)=\lambda \frac{\log_2 N}{N}+(1-\lambda) \left( -1+ 1/2 \log_2 N \right) .
\end{align}

It is interesting to note that the link length distribution of EPT's follows an inverse power law $P(w)\propto w^{-\alpha}$ with $\alpha=2$. This is verified by the observation that for $k=0,...,\log_2 N$ there are $2^{k+1}$ links that establish connections between nodes at level $k$ and nodes at level $k+1$. All of these links have equal length $w=2^{-(k+1)} N$, such that $P(w)\propto w^{-2}$. The relationship $P(w) \propto w^{-2}$ still holds for EPT's with $\kappa>2$, or also for random partitioning trees for which the new root nodes are chosen at random from the interval $[n_1,n_2]$ in step 2 of the construction algorithm. This is seen from the self-similarity of the trees, i.e. when considering length scales of order $w/\kappa$ one essentially considers $\kappa$ copies of the orginal tree, i.e. the link length distribution $P(w)$ of the tree has the scaling property $P(w/\kappa)\propto \kappa^2 P(w)$. The solutions to this scaling relationship are power laws $P(w)\propto w^{-\alpha}$ with $\alpha=2$. 

Note, that above we have always embedded nodes in space in such a way that typical distances were scaled by a factor $1/\kappa$ when proceeding to the next level of the hierarchy and multiplying the number of nodes by a factor of $\kappa$. Following this procedure the embedding of the nodes in space is dense, i.e. the tree includes every node in space. Trees that don't include all nodes in space can be constructed if one introduces a scaling relationship for length scales $\nu$ smaller than $1/\kappa$ when proceeding from one level of the hierarchy to the next, cf. Fig. \ref{F-3}b for an illustration of such a tree for $\kappa=2$ and $\nu=1/3$. 

A simple construction principle for such EPTs with general $\kappa$ and $\nu$ could be the following recursive procedure:
\begin{itemize}
 \item [1.] Select a root node $n_0$ and a length scale $L_0$ and divide the interval $[n_0-L_0/2,n_0+L_0/2]$ into $\kappa$ parts assigning the nodes $n_0-L_0/2+L_0/(2\kappa$), $n_0-L_0/2+3L_0/(2\kappa)$, $...$, $n_0+L_0/2-L_0/(2\kappa)$ the role of root nodes for the next level of the hierarchy. Proceed with step 2 for each of the new root nodes and set the length scale to $L_1=\nu L_0$.
 \item [2.] Stop the procedure if the length scale $L_k$ is smaller than the spacing of the grid. Otherwise, given a root node $n_k$ and a length scale $L_k$  divide the interval $[n_k-L_k/2,n_k+L_k/2]$ into $\kappa$ parts assigning the nodes $n_k-L_k/2+L_k/(2\kappa$), $n_k-L_k/2+3L_k/(2\kappa)$, $...$, $n_k+L_k/2-L_k/(2\kappa)$ the role of root nodes for the next level of the hierarchy. Proceed with step 2 for each of the new root nodes and set the length scale to $L_{k+1}=\nu L_k$.
\end{itemize}

Even though for $\nu<1/\kappa$ the tree contains only part of the nodes, several such trees, possibly with different numbers of hierarchical levels, can be embedded in space and connected by a few links, such that the combined tree essentially still has the same link length distribution as an EPT with $\nu<1/\kappa$. By constriction, the link length distribution of EPT's with general $\kappa$ and $\nu$ obeys the scaling relationship $P(\nu w)\propto \kappa/\nu P(w)$. Assuming a power law ansatz $P(w)\propto w^{-\alpha}$ one obtains
\begin{align}
 \alpha= -\log_\nu \kappa +1,
\end{align}
reproducing the special case of $\alpha=2$ for $\kappa=1/\nu$ that we discussed above.

Since the construction only affects the location of nodes in space, EPT's with $\nu<1/\kappa$ have the same average pathlength as an EPT with $\kappa=1/\nu$. However, since for $\nu<1/\kappa$ one has $\alpha(\kappa,\nu)<\alpha (\kappa,1/\kappa)$ EPT's with $\nu<1/\kappa$ require more wire than EPT's with $\nu=1/\kappa$, i.e. they are suboptimal and one does not expect them to occur when optimizing $E(\lambda)$ at a global level. Nevertheless, they give an interesting reference case for the case of the competitive optimization considered in section \ref{LocalOptima}.

\subsection{Transitions}
\label{sub3}

The expressions in Eq.'s (\ref{Efull}), (\ref{Estar}), (\ref{Ering}) and (\ref{Eept}) already illuminate that a similar transition as the one described in \cite{Sole} between star-like and fully connected networks will occur in spatially constrained networks. Clearly, for $\lambda=0$ one has $E_\textrm{full}<E_\textrm{star}<E_\textrm{EPT}<E_\textrm{ring}$, i.e. the complete graph is optimal.

When $\lambda$ becomes substantially larger than ${\cal O}(1/N)$ complete graphs lose competitiveness since the contribution of the cost of wire scales linearly with $N$. Then, out of our four example networks over a wide range of $\lambda$-values star networks are the preferred configurations. Eventually, when the cost of wire becomes the dominant consideration for $\lambda=1$, one has $E_\textrm{star}=1/2$, $E_\textrm{EPT}\sim \log_2/N$ and $E_\textrm{ring}\sim 1/N$, i.e. ring graphs are optimal. 

EPT's, which are larger than stars, but require less wire for their construction than stars, will become preferred in the regime in between the star network regime and the regime very close to $\lambda=1$ in which the ring graphs are optimal. To see this, consider $\lambda_N=1-1/N$. In leading order one then has $E_\textrm{EPT}(\lambda_N)\propto 1/N \log_2 N$, $E_\textrm{star}(\lambda_N)=1/2$ and $E_\textrm{ring}=1/4$, i.e. EPT's outperform both star networks and ring graphs.

These rough back of the envelope calculations thus sketch out a phase diagram with several possible transitions. For $\lambda\ll 1/N$ optimal networks are fully connected and a transition towards star networks is expected for increasing $\lambda$. Increasing the relative cost of wire further, for $1-\lambda\ll 1/N$ a second transition from star-like networks towards ring graphs is expected.  Our above arguments also show that at the transition, in a small parameter regime close to $\lambda=1$ hierarchical partitioning trees with power law link length distributions become optimal.

Of course, the three discussed network topologies only stand for `typical' network configurations that illustrate that transitions between homogeneous and heterogeneous networks can be expected when increasing $\lambda$, and a second transition from heterogeneous networks towards linear chains can be reckoned with when further increasing $\lambda$ close to 1. To explore the full `phase diagram' between both extremes, we proceed with a numerical simulation to construct networks for $0<\lambda<1$.

\subsection{Numerical results and analysis}
\label{sub4}

In the following construction of optimal networks we employ a numerical optimization scheme that operates via simulated annealing, similar to previous work, cf. \cite{Gopal}. The scheme is typically seeded with a $k$-regular lattice with  small $k$. At each iteration of the scheme a rewired network configuration that differs from the previous configuration either by (i) the addition of a link at a randomly selected link vacancy, (ii) the deletion of a randomly selected link, or (iii) the exchange of the endpoints of two randomly selected pairs of connected nodes is suggested. Then the energies of the original and rewired configurations, $E$ and $E^\prime$, are calculated. The rewired configuration is accepted with probability $q=1$ if $E^\prime<E$ and with probability $q\propto \exp(-r t (E^\prime-E))$ otherwise. In the above $t$ stands for the iteration number and the constant $r$ determines the rate of cooling down in the annealing scheme. In the procedure connectedness of the networks is enforced and the formation of loops is prevented.

It is worthwhile to note that in the above we only fix the system size and the constant $\lambda$. The complete optimal network structure, i.e. the number of links, the amount of wire required, and structural properties of the optimal network topologies `emerge' from the guided rewiring scheme.

\begin{figure*}[tbp]
\begin{center}
\includegraphics [width=.95\textwidth]{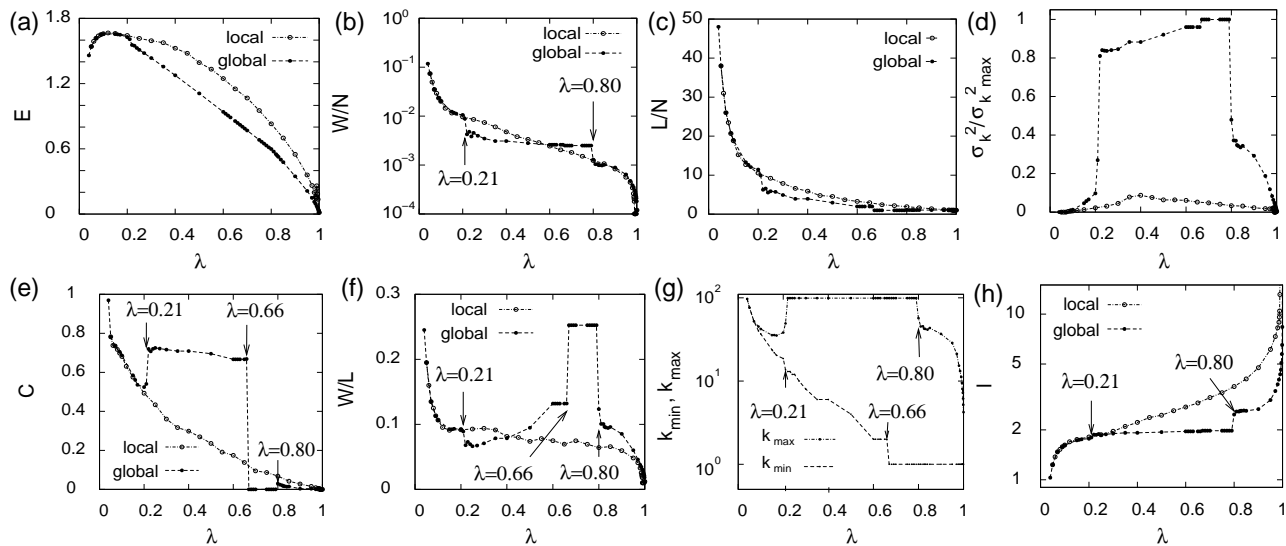}
\caption{Phase diagram showing the dependence of E on $\lambda$ for the global optimum (filled circles) and for the local optimum from a competitive individual optimization process (open circles). The arrow points to critical points which mark the transitions between different types of networks, see text. The panels show (a) the energy $E(\lambda)$, (b) the amount of wire per node, (c) the number of links per node (degree), (d) variation of the degree (normalized by the star configuration), (e) clustering coefficient, (f) average length of a link $W/L$, (g) maximum and minimum degrees for the global optimum configuration and (h) average pathlengths. Data are for $N=100$ and represent averages over at least 100 networks for each value of $\lambda$.}
\label{F-1}
\end{center}
\end{figure*}


To elucidate the dependence of optimal network configurations on the parameter $\lambda$, we constructed ensembles of at least 100 optimal networks of $N=100$ nodes for systematically varied values of $\lambda$. To characterize the optimal topologies and their embedding in space, we measured the energies, cf. Eq. (\ref{E0}), the average amount of wire per node $W/N$, the average degree $k=2L/N$, the degree variation $\sigma^2_\text{k}=\sum_i (k-\langle k\rangle)^2$, the clustering coefficient $C$, the average length of a link $w=W/L$, maximum and minimum degrees $k_\text{max}$ and $k_\text{min}$ and the average shortest pathlength $l$. Figure \ref{F-1}, panels (a)-(h) gives an overview over these network statistics when the parameter $\lambda$ is changed and Fig. \ref{F-2} schematically illustrates typical optimal network configurations for different values of $\lambda$.

 While the dependence of the energy $E(\lambda)$ on $\lambda$ in panel (a) of Fig. \ref{F-1} is smooth, several network characteristics reveal discontinuities. Most prominently, in panel (b) of Fig. \ref{F-1} one notes jumps in the amount of wire per node $w=W/N$ at $\lambda=0.21$ and $\lambda=0.80$. The first point marks a connectivity transition from densely wired networks to sparsely wired star-like networks. Below $\lambda=0.21$ one finds close to regular highly cliquish long range worlds. Networks above $\lambda=0.21$ are characterized by the presence of a single central hub node, that connects to all other nodes. The long range links between the hub and the periphery nodes are complemented by a large number of short range links that connect nearest neighbours in space. Optimal networks are networks composed of a $k$-regular periphery network and an additional central hub node that links to all other nodes. The value of $k$ above gives the number of short range connections of periphery nodes. As $\lambda$ is increased and wire becomes more and more expensive, more and more of these short range links are dropped, which leads to a series of transitions towards lower values of $k$, till finally at $\lambda=0.66$ a transition from the $k=2$ (`wagon wheel') network to a star configuration with $k=0$ occurs, cf. the drop of the clustering coefficient in panel (e) and the steep increase of the average link length in panel (f). The star persists until at $\lambda=0.8$ the long range connections required in the star become too expensive and the central hub is replaced by a collective of hub nodes of roughly similar degree. These hub nodes, which are homogenously distributed in space, serve as local centres and link to nodes closest to them in space. Initially for $\lambda$ close to the transition point $\lambda=0.8$ the hubs are tightly interlinked by long range links, which leads to a non-vanishing clustering coefficient: note the little `hump' at $\lambda=0.8$ in panel (e). Excluding the short cycles between the hub nodes the remainder of the network is tree-like. Increasing $\lambda$ further, long range connections between hub nodes are thinned out and the number of hub nodes increases, while also the difference in the degrees of hub nodes and periphery nodes becomes smaller. Cycles in the networks become increasingly fewer and longer, eventually leading to the appearance of trees. The organization of these trees is hierarchical and very similar to EPT's. When $\lambda$ is extremely close to 1, such that the economy of wire dominates but short pathlengths still give a small contribution to Eq. (\ref{E0}), a ring graph is optimal. Finally, at $\lambda=1$ a linear chain, in which nearest neighbours in space are connected, becomes optimal.

In the light of the scale-free networks found in the sparsely connected regime in \cite{Sole} a closer investigation of the transitions appears of interest. It turns out that the transition from star-like to exponential and linear networks when increasing $\lambda$ does not lead via a regime in which scale-free networks appear. The reason for this is found in the embedding in an underlying space: the central hub of the star networks is replaced by a collective of hubs of equal degrees when increasing $\lambda$ above $\lambda=0.8$. When further increasing $\lambda$ the number of these hub nodes increases, but they are always found to be of roughly similar degree, thus causing a distinct peak in the degree distribution. 

\begin{figure}[tbp]
\begin{center}
\includegraphics [width=.35\textwidth]{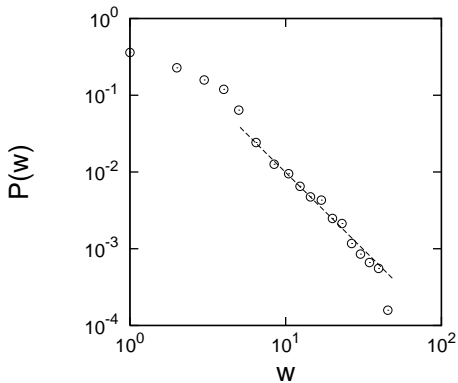}
\caption{Link length distribution of optimal networks of $N=100$ nodes constructed for $\lambda=0.98$. The line indicates a power law $P(w)\propto w^{-\alpha}$ with exponent $\alpha=2$.}
\label{F-1b}
\end{center}
\end{figure}

The sparse regime is also of interest since Ref. \cite{Gopal} or-- in the context of synchronization problems in space-- Ref. \cite{MB2} reported power laws in the link size distribution when wire is expensive. Since the star configurations are associated with uniform link length distributions such power laws can only be found for $\lambda>0.8$. In this regime, however, the networks are essentially tree-like, such that comparisons to real-world networks as those of \cite{Yook,Payman,Schuz,Egui}  appear unrealistic. Nevertheless, for $1-\lambda\ll 1$ the data give some support for power law tails of the link length distributions, cf. Fig. \ref{F-1b} which plots the link length distribution for optimal networks evolved for $\lambda=0.98$.  For all the cases we investigated we found exponents slightly below $\alpha=-2$, thus being much larger than reported in \cite{Gopal}, but in good agreement with the link length distributions expected for EPT's and partitioning trees, cf. subsection \ref{sub2}. The difference between our findings and \cite{Gopal} may be attributable to the variability in the number of links in the problem investigated in our study.


\section{Local Optima}
\label{LocalOptima}

By defining normalized node contributions to wire $\overline{w}_i=1/(N-1)\sum_{j} a_{ij} d(i,j)$ and pathlength $\overline{l}_i=1/(N-1)\sum_j l(i,j)$ a node contribution to the fitness or energy $E(\lambda)$ (cf. Eq. (\ref{E0})) can be introduced
\begin{align}
\label{e0}
 e_i(\lambda)=\lambda \overline{w}_i + (1-\lambda) \overline{l}_i.
\end{align}
Trivially, $E(\lambda)=2/N \sum_i e_i(\lambda)$. However, would a set of nodes where each node individually strives to optimize its own wire-pathlength balance also reach the optimum global configuration that we discussed in the previous section? While this appears possible in the symmetrical regular grid phase, we argue below that it is impossible in the `symmetry-broken' phases where hub and periphery nodes coexist.

To construct networks where nodes optimize their individual wire-communication balances, we consider the following process. After seeding the algorithm with a ring lattice, in each iteration a node, e.g. node $i$, is chosen at random. Then, either a link from this node to a randomly selected neighbour is removed or a link between the node and another randomly selected node is introduced. This link addition/removal process generates a rewired network configuration, which is accepted if it reduces $i$'s  energy balance $e_i(\lambda)$ and rejected otherwise. If the altered configuration is rejected we proceed with the previous configuration. The process is repeated for a sufficiently large number of iterations, such that the average of the number of links (over intermediate timescales) is stationary. In the process above, connectedness of the networks is ensured and the formation of loops is prevented.

\begin{figure*}[tbp]
\begin{center}
\includegraphics [width=.95\textwidth]{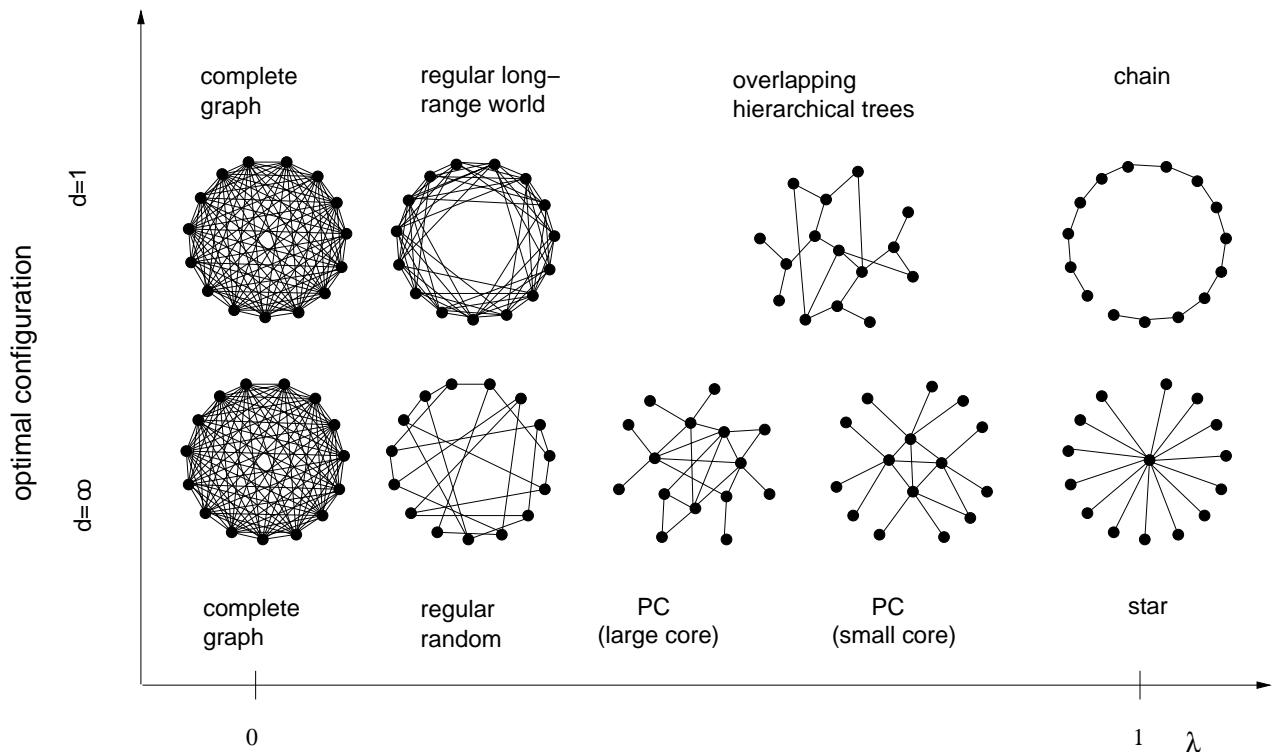}
\caption{Illustration of optimal networks in the order in which they appear when the trade-off parameter is increased from $\lambda=0$ to $\lambda=1$.}
\label{LO}
\end{center}
\end{figure*}

\subsection{The case of $d=\infty$}

Let us first consider the case of an infinite-dimensional underlying space, for which (up to a scaling factor) we set $d(i,j)=1 \forall i,j$, reproducing the problem considered in \cite{Sole}. Thus for $d=\infty$  Eq. (\ref{e0}) reduces to
\begin{align}
 e_i(\lambda)=\lambda k_i/(N-1) + (1-\lambda) \overline{l}_i,
\end{align}
where $k_i$ is the degree of node $i$ and $\overline{l}_i$ the average shortest pathlength from node $i$ to all other nodes as above. To understand the instability of the star configuration in a situation when individual nodes optimize their individual cost of link-communication balances, let us consider the `wagon-wheel' configuration discussed above in section \ref{Global}, cf. Fig. \ref{F-1b}(c). First, consider the central hub node, which has degree $k_\text{hub}=N-1$ and average network distance $\overline{l}_\text{hub}=1$. Thus
\begin{align}
e_\text{hub}=2-\lambda. 
\end{align}
In contrast, a periphery node has $k_\text{per}=3$ and, neglecting ${\cal O} (1/N)$ terms, average pathlength $\overline{l}=2$, which gives
\begin{align}
e_\text{per}=\lambda 3/N+ 2 (1-\lambda).
\end{align}
While benefitting from a slightly improved communication the hub node has to pay for almost all the wire! Thus, in any large system with $N\gg 1$, a hub node would strive to drop links to improve its own cost of links-communication balance. This argument makes it clear that the star configurations that were found to be optimal in a wide range of $\lambda$ parameters in \cite{Sole} can generally not survive in a local optimization process as introduced above.

Before proceeding, let us first briefly recap the dependence of typical network structures on the parameter $\lambda$ as explained in \cite{Sole}. For large  $\lambda$, i.e. expensive links, a sparse regime is found. Initially, for $\lambda$ close to 1 networks are trees with exponential degree distributions. Decreasing $\lambda$ the degree distributions broaden and for some intermediate values of $\lambda$ scale-free trees are observed. At some critical value of $\lambda$ an abrupt transition towards star-like topologies takes place. Finally, decreasing $\lambda$ further a second transition from the star-like networks towards fully connected networks is found.

\begin{figure}[tbp]
\begin{center}
\includegraphics [width=.45\textwidth]{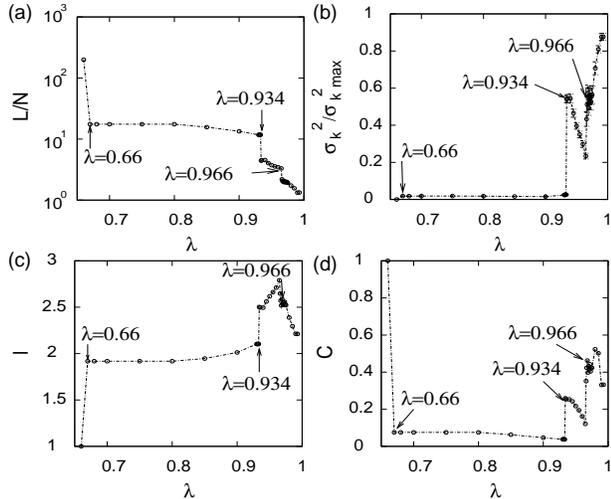}
\caption{Dependence of (a) the number of links per node, (b) the relative degree variance, (c) the average shortest pathlength, and (d) the relative clustering coefficient on the parameter $\lambda$ for $d=\infty$ and $N=100$. Note the transition points at $\lambda=0.66$ and $\lambda=0.934$.}
\label{FH}
\end{center}
\end{figure}

Figure \ref{FH} gives a summary of network statistics that illustrate changes in the structure of the competitively evolved networks when the parameter $\lambda$ is changed, while Fig. \ref{LO} presents a graphical illustration of typical optimal network topologies in the order in which they appear when $\lambda$ is increased from $0$ to $1$. In Fig. \ref{FH}, as a guide for the discussion of the evolved network structures, we consider the number of links per node $L/N$ (panel a), variance of the degree sequence (panel b), average shortest pathlength (panel c) and the average clustering coefficient (panel d). The degree variance has been normalized by the degree variance of a star network. 

As in the globally optimized networks of \cite{Sole}, sharp transitions between different network regimes are evident. The first point of interest is a transition towards fully connected networks at around $\lambda_1=0.66$. A second transition, marked by an abrupt drop in the number of links accompanied by a steep increase in degree variance, distance and clustering coefficient, is found at approximately $\lambda_2=0.934$. A third transition is characterised by another sharp drop in the number of links, sharp increase in the degree variance and clustering, but decrease in the average shortest pathlength at around $\lambda_3=0.966$. Thus, excluding the fully connected topology, three types of network configurations need to be discussed.

In the first non-trivial regime, $\lambda_1<\lambda<\lambda_2$, the networks have distinctly smaller than expected degree variances. Other network statistics reveal no significant deviation to random expectations, the networks thus being very close to random regular graphs. At the first glance these networks appear very similar to the entangled nets discussed in \cite{Donetti}. However, in contrast to the latter they have small, but non-vanishing clustering coefficients.

Above the second transition, $\lambda_2<\lambda<\lambda_3$ the permutation symmetry of the previous regime is broken and the degree distributions become bimodal. The optimal networks become periphery-core networks composed of a core of hub nodes and a periphery of low degree nodes. Such networks have, e.g., been discussed in the context of networks in the immune system, cf. \cite{Behn}, but they also naturally occur when optimizing networks for diameter while requiring robustness against targetted node removals \cite{Pongor}. The periphery-core organization essentially results from the instability of the star configuration in the competitive optimization procedure. When selected for optimization, hub nodes strive to eliminate parts of their connections. However, since being outnumbered by the many periphery nodes, hub nodes are not able to drop connections with the same frequency as other nodes attempt to connect to them. Furthermore, while striving to disconnect from periphery nodes, it is beneficial for hub nodes to maintain connections with other hub nodes since this reduces their respective average pathlengths. These assortative connections between the hub nodes and the desirability for periphery nodes to form connections to hub nodes, lead to  a strongly cliquish network arrangement (cf. panel d). This finding appears of interest since it illustrates a novel mechanism for the appearance of clustering, i.e. via the competitive optimization of cost of links and communication balances of individual nodes. To our knowledge this has not been discussed in the literature before.

Further increasing the cost of wire, the number of links decreases steadily. Since periphery nodes have only few connections to the core, connections between core nodes are thinned out, which reduces their respective degrees and thus decreases the degree variance in the network (cf. panel Fig. \ref{FH}b). As most triangles in the network result from the tightly connected core, thinning out the core also reduces the clustering coefficient and leads to an increase in average shortest pathlength. At $\lambda=0.966$ there is a further transition towards another periphery-core organization, distinguished by a sharply reduced core size. As links becomes more and more expensive, this core gradually reduces in size, a process that further increases the degree heterogeneity, contracts the networks and eventually reduces clustering as the number of hub nodes decreases. Further link cost increases finally drive the networks towards a star configuration.

\subsection{The case of $d=1$}

Consider again the `wagon-wheel' topology introduced in section \ref{Global} and let us compare the hub node and a periphery node farthest away from the hub. Neglecting ${\cal O}(1/N)$ contributions the energy balance for the hub node reads
\begin{align}
 e_\text{hub}=\lambda N/2 + (1-\lambda),
\end{align}
whereas
\begin{align}
 e_\text{per}=2-3/2\lambda
\end{align}
for the periphery nodes farthest away from the hub. In $d=1$ the energy difference between hub and periphery nodes is thus larger than for $d=\infty$ and it increases in proportion to the number of neighbours of the central hub.

Comparing the cases of $d=\infty$ discussed in the previous subsection and the case of $d=1$ of interest here, a relevant consideration is also the `expense' a periphery node would incur to link to a hub node. Whereas for $d=\infty$ this is just $\Delta w=1/(N-1)$ (i.e. one link), in $d=1$ it varies between $\Delta w=1/(N-1)$ for a node that is a spatial nearest neighbour of a hub and $\Delta w=1/2$ for a node at the farthest possible distance from the hub. For an `average' node one has $\Delta w \approx 1/4$. Thus, on average in $d=1$ the cost of hub formation is larger than for $d=\infty$. Benefits from hub formation accrue from reduced pathlength, and are independent of the dimension of the underlying space. It appears a reasonable estimate to use the difference in pathlength that results from directly linking to a hub node and linking to a node which is connected to a hub. Hence reductions in average pathlength from linking to another node are at most of order 1. Since for large systems $\Delta w\sim 1/N<\text{const.}$ for $d=\infty$  and $\langle \Delta w\rangle =\text{const.}$ in $d=1$ this heuristic argument suggests why hub formation is suppressed for $d=1$.

To gain a better understanding of the network evolution process we have analyzed the dependence of typical network configurations on the parameter $\lambda$. Data for some averaged network statistics, compared to the global optimum, are displayed in the panels of Fig. \ref{F-1}, but see also Fig. \ref{LO} for a graphical illustration of the optimal network topologies in $d=1$.  Two main observations stand out. First, panel (a) of Fig. \ref{F-1} clearly illustrates that there is a deviation between the energies $E(\lambda)$ obtained by  global and local competitive optimization for large $\lambda$, while both procedures reach similar energies for small $\lambda$. Closer inspection shows that the point where differences start to appear indeed coincides with $\lambda_\text{c}=0.21$, the transition point at which the globally optimal configuration is realized by a single central hub. This is explained by our above argument that the central hub configuration is not stable in the competitive process. Second, all other network statistics show the absence of sharp transitions in the competitive optimization when varying $\lambda$. This is again not surprising, as the global optima characterized by the existence of hub nodes are unstable in the competitive process. Rather, the competitive optimization generates densely connected regular graphs for low cost of wire and then gradually increases the variance in degrees as wire becomes more expensive, cf. panel (d). As this happens, the amount of wire and number of links decrease gradually (panels (b) and (c)), also entailing gradual decreases in the average link length (panel (f)), clustering coefficient (panel (e)) and average pathlength (panel (h)).

\begin{figure}[tbp]
\begin{center}
\includegraphics [width=.45\textwidth]{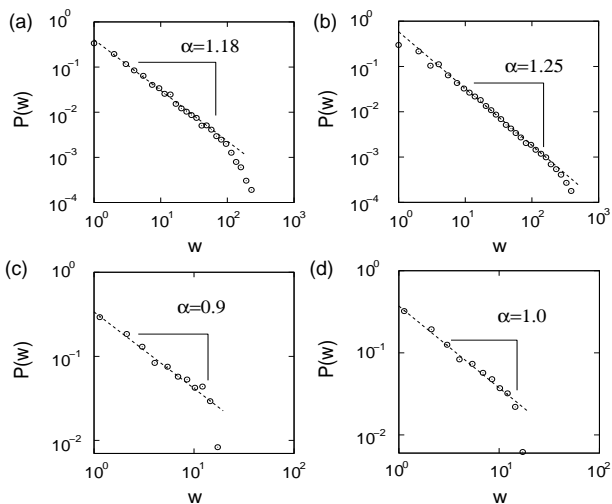}

\caption{Distribution of link lengths $P(w)$ for a system of $N=900$. Distributions are for $d=1$ and $\lambda=0.95$ (a), $d=1$ and $\lambda=0.5$ (b), $d=2$ and $\lambda=0.95$ (c), and $d=2$ and $\lambda=0.9$ (d). The straight lines indicate power laws $P(w)\propto w^{-\alpha}$ with exponents $\alpha=1.18$, $\alpha=1.25$, $\alpha=0.9$ and $\alpha=1.0$, respectively. The data represent averages over 5 network configurations and are binned logarithmically. Note that for $N=900$ the maximum distance on the lattice is $d_\text{max}=450$ for $d=1$ and $d_\text{max}\approx 21.2$ for $d=2$.}
\label{F2}
\end{center}
\end{figure}

The suppression of hub formation for expensive wire has an important consequence: it also impedes the development of uniform length distributions associated with the star configurations. As a result, the power laws in the link length distributions (observed for very expensive wire in section \ref{Global}) persist over a much wider range of $\lambda$ parameters. Panels (a) and (b) of Fig. \ref{F2} illustrate the emergence of the power law behaviour for a large system of $N=900$ nodes. Over several orders of magnitude the tails of the link size distributions are very well described by power laws $P(w)\propto w^{-\alpha}$ with exponents $\alpha=1.18$ (for very expensive wire $\lambda=0.95$) and $\alpha=1.25$ (for intermediate cost of wire $\lambda=0.5$). In contrast to the observations of Ref. \cite{Gopal}, which reported exponents $\alpha>2$, the competitive optimization thus also leads to a much less steep decay of link lengths. These results are also consistent with \cite{Petermann}, who find that a threshold exponent $\alpha_\text{c}=d+1$ in the power laws of link length distributions separates a small world from a long world regime. Applying the results of \cite{Petermann} to the case of $d=1$, small worlds require $\alpha<2$, which is what we have observed in our simulations.

The deviation of the power law exponents from $\alpha=2$ is also quite interesting when compared to the hierarchical partitioning trees introduced in subsection \ref{sub2}. Since exponents $\alpha<2$ naturally occur for the (globally) suboptimal trees with $\nu<1/\kappa$ one may argue that the structure of the competitively evolved networks comprises multiple overlapping suboptimal EPT's. Changing the cost-pressure of the wire $\lambda$ would smoothly lead to different spatial arrangements of the EPT's, which agrees with the smooth increase of the power law exponent $\alpha$ observed in the numerical simulations when $\lambda$ is increased.

\subsection{The case of $d=2$}

In this section we briefly comment on the link length distributions under competitive optimization in $d=2$, which appears the most relevant case for many practical applications.

Let us again follow the heuristics developed in the previous subsection and consider the costs and benefits of hub formation. To link to a hub node, wiring costs $\Delta w$ in $d=2$ range between $1/(N-1)$ and $d_\text{max}(2)/(N-1)$ and are, on average, of order $\Delta w\sim N^{-1/2}$. Thus, for a large enough system we observe that $\Delta w<\text{const}$. In fact, extending the above argument to arbitrary $d$, one has $\Delta w\sim N^{1/d-1}$, suggesting that hub formation becomes easier in higher dimensions.

As already observed in the introduction, cases of $d>1$ are hard to access in numerical simulations, since the maximum distances on higher dimensional lattices decrease strongly with the dimension for constant system size. For this reason, it is not possible to obtain reliable numerical results for the full phase diagram for a reasonable system size. We have, however, observed hub formation in $d=2$ for $N=900$ in numerical simulations for $\lambda=0.99$. For smaller $\lambda$, when hub formation is still suppressed, the power law behaviour in the link length distributions observed for $d=1$ is recovered. Figure \ref{F2} (c) and (d) display data for two example situations. All cases we have investigated suggest a more gradual decay of the link length distributions in $d=2$ than in $d=1$, measured exponents being in the range $\alpha \approx 0.9 ... 1.0$.

\section{Summary and discussion}
In this paper we have considered networks that optimize a balance between the (infrastructure-) cost of links and a measure for communication, i.e. average shortest pathlengths. In contrast to previous studies like \cite{Gopal,Sole}, the number of links was not fixed and networks were embedded in space. Additionally, the analysis included not only the globally optimal solutions, but also competitive solutions, where individual nodes strive to optimize their local cost-communication balances. Our analysis reveals marked differences between both situations and we have analysed and discussed how these differences depend on the dimensionality of the underlying space.

Depending on the cost of wire, in globally optimal networks three phases can be distinguished: (i) fully connected graphs, (ii) (almost) regular small worlds, and (iii) star-like networks. Even though the location of the transitions between these phases depends on the dimension, this general structure of the phase diagram holds for all $d$. In a small interval of $\lambda$-parameters shortly before chains become optimal, tree-like networks with power law link length distributions were found. The structure of these trees is hierarchical and the exponent of the power laws in the link length distributions agrees well with results derived for hierarchical partitioning trees.

In the competitive optimization process only the network structures that occur in phases (i) and (ii) were found to be stable and the structure of the phase diagram was found to change when changing the dimension of the underlying space.

For $d=\infty$, increasing the cost of wire transitions between a complete graph, regular small worlds and different types of periphery-core networks have been found. Hence, in the competitive optimization process, the regime of star-like networks discussed in \cite{Sole} is replaced by regimes of networks composed of a highly cliquish core consisting of hub nodes and a periphery composed of low degree nodes. When varying the cost of wire, sharp transitions occur between different periphery-core arrangements, characterized by different core sizes.

For $d=1$ hub-formation is suppressed in the competitive optimization and thus sharp transitions between different regimes are absent. In spite of this, we find power law link length distributions with exponents larger than observed for very expensive wire in the globally optimal networks. The parameter regime where such power laws persist is much larger than for the globally optimal networks. Optimal networks in this regime are also not tree-like, but relatively densely connected. Power law link length distributions with exponents $\alpha<2$ hint at the presence of overlapping suboptimal hierarchical trees.

We have argued that hub-formation again occurs for $d>1$. The regime of trade-off parameters where star-like networks dominate grows with the dimension of the underlying space. Nevertheless, power laws in the link length distribution could be recovered in the regime (ii) of (almost) regular  small worlds. The observed exponents were found to be in the range of exponents recently observed for systems embedded in 2-dimensional space, e.g., for the internet \cite{Yook} and for integrated circuits \cite{Payman}.

As a last point, we observe that the regime of power law link length distributions is bounded in two regards. The first limitation is when the cost of wire becomes very cheap. In this case link length distributions become uniform at longer length scales, while reflecting the details of the underlying space at short length scales. The second bound is hub-formation. Even though, in principle, hubs are unstable in the competitive optimization process, the formation of hubs can occur when a majority of other nodes attempt to link to prospective hubs faster than they can drop links when it is their turn to optimize their position in the network. This process results in the formation of periphery-core networks. Comparisons of the costs and benefits of hub formation indicate that it becomes more prevalent when increasing the dimension of the underlying space.


\acknowledgements

This research was undertaken on the NCI National Facility in Canberra, Australia, which is supported by the Australian Commonwealth Government. I wish to thank F. Boschetti and Nicky Grigg for comments about this work and one anonymous reviewer for suggestions that helped to improve the manuscript.

\appendix*

\section{Calculation of the average pathlength for EPT's}
\label{appendix}
To calculate the average pathlength for EPT's consider an EPT comprised of $N=2^{k+2}-1$ nodes with root node $O$ and denote the two branches emerging from the root node by $A$ and $B$. Correspondingly, denote the root nodes of the branches by $O_A$ and $O_B$. Since $N=2^{k+2}-1$ the whole tree consists of $k+1$ hierarchical levels, whereas the subtrees $A$ and $B$ have $k$ levels each and are both comprised of $N_A=N_B=2^{k+1}-1$ nodes.

We first consider the length of a path from a node at level $k_A$ in subtree $A$ to a node at level $k_B$ in subtree $B$. From level $k_A$ one needs $k_A$ steps to reach $O_A$, $2$ steps to get to branch $B$ via $O_A$ and $O$ and a further $k_B$ steps to reach level $k_B$ in subtree $B$. Hence, the average pathlength from a node at level $k_A$ in subtree $A$ to nodes in $B$ is
\begin{align}
 l_{k_A}&=\frac{1}{2^{k+1}-1} \sum_{k_B=0}^k (2+k_A+k_B) 2^{k_B}\\
        &=(2+k_A)+\frac{(k-1)2^{k+1}+2}{2^{k+1}-1}.
\end{align}

This allows to calculate the average pathlength between nodes in $A$ and nodes in $B$, which is
\begin{align}
 l_{AB}&=\frac{1}{2^{k+1}-1} \sum_{k_A=0}^k 2^{k_A} l_{k_A}\\
       &=2+2 \frac{(k-1)2^{k+1}+2}{2^{k+1}-1}.
\end{align}

One notes that $l_{AB}\approx 2k$ when $2^{k+1}\gg 1$. Similarly, to reach a node at level $k_A$ in subtree $A$ from $O$ $k_A+1$ steps are required. Thus, the average pathlength from $O$ to $A$ is
\begin{align}
 l_{OA}&=\frac{1}{2^{k+1}-1} \sum_{k_A=0}^k (1+k_A) 2^{k_A}\\
       &=l_{AB}-1.
\end{align}

Making use of the symmetrical structure of the tree for the average pathlength of the EPT one thus has
\begin{align}
 \overline{l}_\textrm{EPT}&=\frac{2}{N(N-1)} \left( 2N_A l_{OA} + \frac{N_A(N_A-1)}{2} l_{AB}\right)\\
  &=\frac{N_A+3}{4N_A+2}l_{AB}-\frac{2}{2N_A+1}.
\end{align}

For large trees $N\gg 1$ the above simplifies to
\begin{align}
 \overline{l}_\textrm{EPT}= \frac{1}{2} \log_2 N -1.
\end{align}

\end{document}